\documentclass[aps,twocolumn,reprint,amsmath,amssymb,showpacs,superscriptaddress]{revtex4-1}
\usepackage{multirow}
\usepackage{mathtools}
\usepackage{graphicx,epsfig}
\usepackage{epstopdf}
\usepackage{wasysym}
\usepackage{stackengine}
\usepackage{color,soul} 
\usepackage{soul}
\usepackage[colorlinks=true,citecolor=blue,linkcolor=red,linktocpage=true,pagebackref=false]{hyperref}
\usepackage{cancel}
\usepackage[normalem]{ulem}
\usepackage{multibib}

\begin{document}
\renewcommand{\thefootnote}{\arabic{footnote}} 
\newcommand{\be}{\begin{equation}}
\newcommand{\ee}{\end{equation}}
\newcommand{\ba}{\begin{eqnarray}}
\newcommand{\ea}{  \end{eqnarray}}
\newcommand{\ve}{\varepsilon}
\newcommand{\Tau}{\mathcal{T}}
\newcommand{\cals}{\mathcal{S}}
\newcommand{\calz}{\mathcal{Z}}
\newcommand{\calc}{\mathcal{C}}
\newcommand{\cale}{\mathcal{E}}
\newcommand{\cald}{\mathcal{D}}
\title{Slave-Spin 1 formulation: A simple approach to 
time-dependent transport through an interacting two level system. }

\author{Mar\'{\i}a Florencia Ludovico}
\affiliation{Consiglio Nazionale delle Ricerche, Istituto Officina dei Materiali (IOM) and Scuola Internazionale Superiore di Studi Avanzati (SISSA),
Via Bonomea 265, I-34136, Trieste, Italy}
\author{Massimo Capone}
\affiliation{Consiglio Nazionale delle Ricerche, Istituto Officina dei Materiali (IOM) and Scuola Internazionale Superiore di Studi Avanzati (SISSA),
Via Bonomea 265, I-34136, Trieste, Italy}

\begin{abstract}
We introduce and develop a slave-spin mean-field technique for describing a generic interacting two level systems under time-dependent drivings, where an auxiliary $S=1$ spin is added to describe the localized character of the electrons.  
We show that the approach efficiently captures the main effects of the strong correlations as well as the dynamical nature of the driving, while remaining simple enough to allow for an analytical treatment. Our formalism provides a flexible solution method, which can be applied to different device configurations at an extremely small numerical cost. Furthermore, it leads to a very practical description of adiabatically driven systems in terms of frozen static solutions.
\end{abstract}

\maketitle

\begin{figure}[h]
 \includegraphics[width=0.4\textwidth]{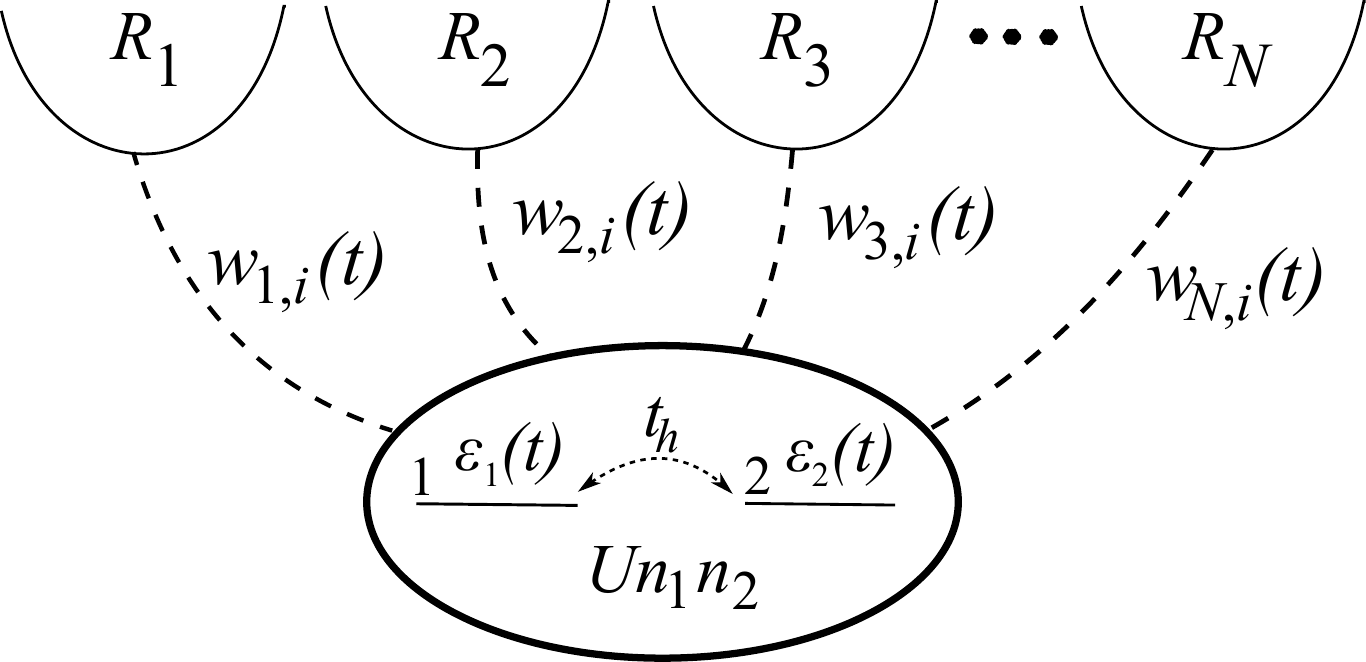}
  \caption{Scheme of the theoretical model considered in this work. It consists of a two energy level system which is in contact to an arbitrary number $N$ of non-interacting electron reservoirs with tunneling amplitudes $w_{j,i}(t)$. Here $i=1,2$ denotes the level number and $j=1,2,...,N$ corresponds to a reservoir index. The two levels, $\varepsilon_1(t)$ and $\varepsilon_2(t)$ are coupled through a Coulomb interaction with and energy $U$. For generality, hopping between the levels with an amplitude $t_h$ is also allowed. All the tunneling rates with the reservoirs and the energy levels can be time-dependent functions. 
  }\label{fig1}
\end{figure}

{\it{Introduction}}: A two-level quantum system in contact with electron reservoirs is one of the most basic but at the same time meaningful theoretical setup for studying the effect of strong correlations on the transport properties of numerous nanoscale devices. We consider a general and simple system 
as illustrated in Fig. \ref{fig1}, where the coupling between two energy levels is dominated by the Coulomb repulsion between electrons, $U n_1n_2$ that depends on the level's occupation number $n_i$. The two levels can be seen as the two spin levels of a magnetic impurity, or two orbitals of an atom or a molecule, in which case the Coulomb repulsion is  a local quantity, but also two spinless single-level quantum dots coupled by a non-local Coulomb repulsion.
These configuration are paradigmatic realizations of  fundamental collective phenomena ranging from the celebrated Kondo peak in the conductance\cite{kondo} of the Anderson impurity model (AIM) to the Coulomb drag observed in  Coulomb-coupled quantum dots\cite{dragor, dragreview}.
The latter effect can be exploited for new technological applications, such as, the implementation of a self-contained quantum refrigerator \cite{dragsa} or a heat diode \cite{dragst5} leading to renewed interest in the subject\cite{dragst1, dragst2, dragst3, dragst4, dragst6,dragse1, dragse2}.

Exploring the above effects when the system is in addition driven by time-dependent on-site energies $\varepsilon_i(t)$ and/or tunneling barriers $w_{ji}(t)$, is still an open field of great importance for diverse areas including those of thermoelectrics \cite{ludovico1, thierschmann, erdman, lim, brandner}, energy harvesting \cite{thierschmann2, ludovico2}, and also quantum optics \cite{qo1, qo2}. Despite the simplicity of a two-level device and that some approaches have been developed to address similar problems \cite{romero, anders, schiro, schiro2}, those effects remain less studied in the presence of an external driving due to the challenges and numerical costs implied by the theoretical descriptions. 
A simple and effective semianalytical framework to explore the dynamics of interacting systems is given by non-equilibrium extensions of slave-particles techniques, like slave-boson\cite{coleman, KR} and more recent slave-spin\cite{demedici, huber, df,caponedemedici} approaches. 
In particular, the $U\rightarrow{\infty}$ Coleman slave-boson approach within the mean-field approximation \cite{citro, ludovico2} and beyond \cite{dong}, and a nonequilibrium slave-spin $1/2$\cite{daniele} have been applied to a single magnetic impurity. 

Here we introduce a time-dependent mean-field slave-spin 1 approach (S-S1), that presents several advantages over other slave-particles methods, namely {\it{i}}) It can be used to describe a generic Coulomb-coupled two-level device like those in Fig. \ref{fig1}, {\it{ii}}) it leads to a reduction of the numerical costs due to a much lower number of parameters describing the Coulomb interaction, and finally {\it{iii}}) it offers a pragmatical and simple way of studying adiabatically driven systems, for which we show that the full dynamics is described in terms of frozen static solutions at every instant of time.

{\it{Model Hamiltonian and Slave-Spin 1 approach:}}
We describe the full system in Fig. \ref{fig1} by the Hamiltonian $H_{FS}(t)=\mathcal{H}(t)+H_R$, where
\ba\label{ham}
\mathcal{H}(t)&=&\sum_{\mathclap{i=1,2}}\Big(\ve_i(t){n}_i+\sum_{\mathclap{\alpha,k_{\alpha}}}w_{\alpha,i}(t)(c^\dagger_{k_{\alpha}}d_i +d_i^\dagger c_{k_{\alpha}})\Big)\nonumber\\
&&+\,t_h\,(d^\dagger_1d_2+d^\dagger_2d_1)+U{n}_1{n}_2
\ea
represents the two-level subsystem along with the tunneling contacts. In the above equation, the occupation operator reads ${n}_i=d^\dagger_id_i$, and $\alpha$ runs over all the reservoirs. Moreover, for completeness we consider a hopping amplitude $t_h$ between the levels or sites. The operator $c^{\dagger}_{k_{\alpha}}(c_{k_{\alpha}})$ belongs to the reservoir denoted by $\alpha$ and creates (destroys) an electron with momentum $k_{\alpha}$. Our results can be applied to an arbitrary choice of $H_R$ describing the non-interacting reservoirs, which are assumed to be at equilibrium. 

We consider a system in which each of the levels can be at most single occupied, so that there are only four possible electronic configurations $(n_1,n_2)$ for the two energy levels: $\mathcal{F}=\{\vert e\rangle=(0,0); \vert s_+\rangle=(1,0); \vert s_-\rangle=(0,1); \vert d \rangle=(1,1)\}$. Within the S-S1 approach, all these configurations are represented by introducing a single $S=1$ auxiliary spin in correspondence with the total fermionic number. This auxiliary spin, like in other slave-spin methods, is not related with any physical magnetic moment or total spin of the two-level subsystem but it is merely a variable having the commutation relations of a $S=1$ spin. In this representation, the physical Fock space $\mathcal{F}$ is mapped onto a larger one $\mathcal{F}^*$ including the auxiliary spin and two fermionic degrees of freedom connected to the physical fermions, which lead to the above four charge states $\vert m^*\rangle$ with $m=e, s_{\pm},d$, plus the auxiliary spin. Then, we associate to each of the real states in $\mathcal{F}$ one of the states living in $\mathcal{F}^*$ in the following way:
\be
\begin{matrix}
  \vert e\rangle \Leftrightarrow \vert e^* ,S_z=-1\rangle; & \vert s_+\rangle\Leftrightarrow \vert s_+^*, S_z=0\rangle\\
  \vert s_-\rangle\Leftrightarrow \vert s_-^*, S_z=0\rangle; & \vert d\rangle\Leftrightarrow \vert d^*, S_z=1\rangle. 
\end{matrix} 
\ee
While the eight remaining states in $\mathcal{F}^*$, as for example $\vert d^*, S_z=0\rangle$ or $ \vert s_+^*, S_z=1\rangle$, are interpreted as unphysical states and they are excluded by enforcing a constraint on the total number of electrons
\be\label{const}
n^*_\Tau=\frac{S_z}{\hbar}+1,
\ee
where $n^*_\Tau=n^*_1+n^*_2$, with $n^*_i={d^*_i}^{\dagger}d^*_i$ and $d^*_i$ being the fermionic operators in the enlarged space. 
Now, the original Hamiltonian in Eq. (\ref{ham}) that contains the Coulomb interaction term, must be mapped onto an auxiliary $\mathcal{H}^*(t)$ acting in the enlarged $\mathcal{F}^*$. For this, we can see that the local operators are equally represented under the transformations: $d_i\rightarrow d^*_iS^-/(\hbar\sqrt{2})$ and $n_i\rightarrow n^*_i$, while the density-density interaction $n_1n_2\rightarrow S_z(S_z+\hbar)/(2\hbar^2)$ can be rewritten in terms of the spin solely. Then,  
\ba\label{ham2}
\mathcal{H}^*(t)&=&\sum_{\mathclap{i=1,2}}\Big(\ve^*_i(t){n}^*_i+\sum_{\mathclap{\alpha,k_{\alpha}}}\frac{w_{\alpha,i}(t)}{\hbar\sqrt{2}}S^-c^\dagger_{k_{\alpha}}d^*_i+\text{H.c.}\Big)\nonumber\\
&&+t_h\left({d^*_1}^\dagger d^*_2+{d^*_2}^\dagger d^*_1\right)\\
&&+\left(\frac{U}{2\hbar}S_z-\lambda(t)\right)\left(\frac{S_z}{\hbar}+1\right),\nonumber
\ea
where $\ve^*_i(t)=\ve_i(t)+\lambda(t)$ are the renormalized energy levels, and $\lambda(t)$ is the Lagrange multiplier enforcing the constraint in Eq. (\ref{const}) at every time. On the other hand, the hopping contribution between the levels remains unchanged since it is represented in the same way as in the original Fock space, i.e. $d_1^\dagger d_2\rightarrow {d_1^*}^{\dagger} d^*_2$.

{\it{Mean-field dynamics:}} So far we have just introduced a new-look representation for the original interacting Hamiltonian ${\cal{H}}(t)$ in an enlarged Hilbert space. As customary in other slave-particles methods, we are going to treat the problem within a mean-field approximation (MFA) that mainly consists of two steps: {\it{i}}) Decoupling fermionic ($f$) and spin ($S$) degrees of freedom, so that all the states in ${\cal{F}}^*$ are factorized as $\vert\psi\rangle=\vert f\rangle\otimes\vert S\rangle$; and {\it{ii}}) Treating the constraint in Eq. (\ref{const}) on average. These assumptions are justified for $U\gg \{\gamma_\alpha, {\bf{V}}_i, \dot{\bf{V}}_i\}$, where $\gamma_\alpha$ is the hybridization with the reservoir $\alpha$ and ${\bf{V}}_i(t)=\left(\varepsilon_i(t),w_{1i}(t),...,w_{Ni}(t)\right)$ a vector containing all the time-dependent parameters acting on the {\it{i}}-level. We ensure thereby that fluctuations of the spin with respect to the mean values can be neglected, even under the action of slow time-dependent drivings.

Then, step {\it{i}}) of the MFA leads to a noninteracting theory for the fermions with an effective Hamiltonian
\ba\label{hamef}
\mathcal{H}^*_{f}(t)&=&\sum_{\mathclap{i=1,2}}\Big(\ve^*_i(t){n}^*_i+\sum_{\mathclap{\alpha,k_{\alpha}}}w^*_{\alpha,i}(t)c^\dagger_{k_{\alpha}}d^*_i +\text{H.c.}\Big)\\
&&+t_h\left({d^*_1}^\dagger d^*_2+{d^*_2}^\dagger d^*_1\right)+\beta(t),\nonumber
\ea
where $w^*_{\alpha,i}(t)=w_{\alpha,i}(t)\langle S^-\rangle_s(t) /(\hbar\sqrt{2})$ are the renormalized tunneling factors, and $\beta (t)=\frac{U}{2\hbar^2}\left(\langle S_z^2\rangle_s(t)+ \hbar\langle S_z\rangle_s(t)\right)-\frac{\lambda(t)}{\hbar}\left(\langle S_z\rangle_s(t)+\hbar\right)$. Here, the subscript $s$ denotes the part $\vert S\rangle$ of the compound state which corresponds solely to the spin. On the other hand, {\it{ii}}) establishes that the $\hat{z}$ component of the spin evolves according to the constraint
\be\label{const2}
\langle S_z\rangle_s=\hbar\Big(\langle n_\Tau^*\rangle_f-1\Big),
\ee
where, similarly as before, index $f$ means the fermonic state $\vert f\rangle$.
Therefore, this approach steers to a coupled problem between fermionic and spin dynamics, since all the parameters entering Eq. (\ref{hamef}) depend on the spin values which are at the same time determined by the evolution of the fermionic subsystem. 
The evolution of other expectation values or components of the spin can be computed from the equation of motion $-i\hbar d_t\langle {\mathcal{O}}_S\rangle_s=\langle\psi\vert\left[{\mathcal{H}}^*,{\mathcal{O}}_S\right]\vert\psi\rangle$ with ${\mathcal{O}}_S$ being any spin operator. In the case of the raising operator $S^+$ (or equivalently $S^-$) renormalizing the coupling with the reservoirs, it reads
\ba\label{eq1}
-i\hbar d_t\langle S^+\rangle_s&=&(\frac{U}{2}-\lambda)\langle S^+\rangle_s+\frac{U}{2\hbar}\langle\{S_z,S^+\}\rangle_s\nonumber\\
&&-2\langle S_z\rangle_s\sum_{{i=1}}^2\sum_{{\alpha, k_{\alpha}}}\frac{w_{\alpha,i}}{\sqrt{2}}\langle c^\dagger_{k_{\alpha}}d^*_i\rangle_f,
\ea
that depends also on $\langle\{S_z,S^+\}\rangle_s$, for which we have
\ba\label{eq2}
-i\hbar d_t\langle \left\{S_z,S^+\right\}\rangle_s&=&(\frac{U}{2}-\lambda)\langle \left\{S_z,S^+\right\}\rangle_s+\frac{U}{2}\hbar\langle S^+\rangle_s\nonumber\\
&&-2\sum_{{i=1}}^2\sum_{{\alpha, k_{\alpha}}}\frac{w_{\alpha,i}}{\sqrt{2}}\left[\langle {d^*_i}^\dagger c_{k_{\alpha}}\rangle_f\langle {S^+}^2\rangle_s\right.\nonumber\\
&&\left.-\langle c^\dagger_{k_{\alpha}}d^*_i\rangle_f\big(2\hbar^2-3\langle S_z^2\rangle_s\big)\right].
\ea
Now, due to the fact that the most general normalized spin state can be written as \cite{footnote} 
\be
\!\!\!\vert S\rangle\!=\! \sqrt{1\!-\!(\vert d\vert^2\!+\vert e\vert^2)}\vert S_z\!=\!0\rangle+d\vert S_z\!\!=\!1\rangle+e\vert S_z\!\!=\!-1\rangle,
\ee 
all the expectation values for the spin can be expressed as a function of $d$ and $e$, $\langle {\mathcal{O}}_S\rangle_s=g_{{\mathcal{O}}_S}(d,e)$, where $d$ and $e$ play the role of the amplitudes to have double or zero occupancy (see Supplemental Material \cite{supmat} for the specific expressions of $g_{{\mathcal{O}}_S}$). 
Therefore, the coefficients $d(t)$ and $e(t)$ and the Lagrange multiplier $\lambda(t)$ constitute the set of time-dependent variables encoding the full solution of the problem. Their dynamics is obtained solving the system of ordinary differential equations (SODE) composed by (\ref{const2}), (\ref{eq1}) and (\ref{eq2}).

{\it{Stationary case}}: In order to benchmark our approach we start from the case where all the parameters  ${\bf{V}}_i^0=\left(\varepsilon_i,w_{1i},...,w_{Ni}\right)$ with $i=1,2$, are constant functions. For this static configuration, transport through the system could be driven only by the application of bias voltages or temperature differences between the reservoirs. In particular, we are interested in the steady state regime for which all the parameters entering $\mathcal{H}^*_f$ in Eq. (\ref{hamef}) have already attained their stationary value. Thus, we move to a {\it{stationary}} S-S1 formulation, in the sense that the system does not evolve in time. Imposing the stability condition $d_t\langle\mathcal{O}_S\rangle_s^0=0$ for the spin values on Eqs. (\ref{eq1}) and (\ref{eq2}), we find out that in this specific case the spin problem is fully determined by only one real variable, the z-component of the spin $\langle S_z\rangle_s^0=\hbar\left(\vert d \vert^2-\vert e\vert^2\right)$. Any other expectation value of the spin is therefore written in terms of $\langle S_z\rangle_s^0$. Particularly, for the Hamiltonian parameters we have $\vert\langle S^+\rangle_s^0\vert^2=\hbar^2-{\langle S_z\rangle_s^0}^2$ and $\langle S_z^2\rangle_s^0=({\langle S_z\rangle_s^0}^2+\hbar^2)/2$. 
Moreover, Eqs. (\ref{eq1}) and (\ref{eq2}) lead to 
\ba\label{consteq}
0&=&\left[\lambda^0-\frac{U}{2}\left(1+\frac{\langle S_z\rangle_s^0}{\hbar}\right)\right]\left(1-\frac{{\langle S_z\rangle_s^0}^2}{\hbar^2}\right)\\
&&+\frac{\langle S_z\rangle_s^0}{\hbar}\sum_{\mathclap{\alpha}}\!\int\!\!\frac{d\varepsilon}{\pi}\!\mbox{Tr}\!\!\left[\hat{\rho}^0_\alpha(\varepsilon)(\varepsilon-\hat{{H}}^*_0)\right]\!f_\alpha(\varepsilon),\nonumber
\ea
while the constraint reads 
\be\label{eq12eq}
\frac{\langle S_z\rangle_s^0}{\hbar}+1=\sum_\alpha\!\int\!\frac{d\varepsilon}{2\pi}\mbox{Tr}\left[\hat{\rho}_\alpha^0(\varepsilon)\right]f_\alpha(\varepsilon).
\ee
Here the matrix $\hat{\rho}^0_\alpha(\varepsilon)$ is the partial density of state of the two-level system and $[\hat{{H}}^*_0]_{ij}=\varepsilon^*_i\delta_{i,j}+t_h\left(\delta_{i,j+1}+\delta_{i,j-1}\right)$.  $f_\alpha(\varepsilon)$ is the Fermi-Dirac distribution of the reservoir $\alpha$. Details can be found in  \cite{supmat}. 

In this way, the {\it{stationary}} S-S1 allows to describe the effect of a finite interaction $U$ on a generic two-level setup in terms of only two parameters, $\lambda^0$ and $\langle S_z\rangle^0_s$, which are the solutions of the reduced $2\times2$ system of nonlinear equations composed by (\ref{consteq}) and (\ref{eq12eq}) (SNLE). The low dimensionality of the involved system of equations represents one of the advantages of this method with respect to other slave-particles techniques. For instance, the Kotliar-Ruckenstein approach (KR)\cite{KR} for the AIM makes use of seven parameters in total (four bosons and three Lagrange multipliers) for the non-degenerated case. For an explicit comparison with the KR predictions we refer the reader to the Supplemental Material \cite{supmat}.

{\it{Adiabatically driven systems}}: In this stage, we consider an adiabatic driving ($ad$), namely a slow evolution in time of  all the parameters, $\dot{{\bf{V}}}_i\rightarrow 0$. As a consequence of the quasi-static evolution of the system we expect the solutions to remain close to the static solutions at every instant of time $t$. For this reason, the spin values and the Lagrange multiplier may be approximated as\cite{ludovico2} 
\ba\label{solutions}
\langle \mathcal{O}_S\rangle_s^{ad}(t)&\sim & \mathcal{O}_S^t+\Delta\mathcal{O}_S(t),\\
\lambda^{ad}(t)&\sim &\lambda^t+\Delta\lambda(t),\nonumber
\ea
where $\mathcal{O}_S^t\!\equiv\! \langle \mathcal{O}_S^0\rangle_s({\bf{V}}_i(t))$ and $\lambda^t\!\equiv\!\lambda^0({\bf{V}}_i(t))$ are the static values of the observables computed using the values of the parameters ${\bf{V}}_i(t)$ at a given time. Hence, $t$ is used as an index to stress that the dependence on time is purely parametric as in a series of snapshots with frozen parameters.
The first-order corrections $\Delta\mathcal{O}_S(t),\, \Delta\lambda(t)\propto \dot{{\bf{V}}}_i$ take into account the effect of the slow driving. 

Following Refs. \cite{ludovico1, ludovico2}, we can evaluate the SODE in linear response in the small $\dot{{\bf{V}}}_i$ \cite{supmat}. As a consequence of this adiabatic expansion, we find again that the spin values are expressed in terms of $\langle S_z\rangle_s^{ad}(t)$ solely: $\vert\langle S^+\rangle_s^{ad}\vert^2=\hbar^2-{\langle S_z\rangle_s^{ad}}^2$ and $\langle S_z^2\rangle_s^{ad}=({\langle S_z\rangle_s^{ad}}^2+\hbar^2)/2$, as expected  
for a quasi-static evolution. Therefore, $\langle S_z\rangle_s^{ad}(t)$ together with $\lambda^{ad}(t)$ constitute the full set of variables describing the interactions under an adiabatic driving, whose static (or frozen) values $S^t_z$ and $\lambda^t$ are obtained by solving the stationary SNLE with the instantaneous values of the parameters ${\bf{V}}_i(t)$. Moreover, the linear response treatment allows to easily compute the corrections, collected in ${\bf{\Delta}}=(\Delta\lambda, \Delta S_z/\hbar)$, as solutions of a system of linear  equations $\hat{L}(t){\bf{\Delta}}(t)={\bf{C}}(t)$ (SLE), thus
\ba\label{cor}
{\bf{\Delta}}_i=\frac{{\bf{C}}_i\hat{L}_{jj}-{\bf{C}}_{j}\hat{L}_{ij}}{\mbox{det}[\hat{L}]}\,\,\,\,\text{for} \,\,i=1,2\,\,\text{and}\,\,j\neq i.
\ea
The linear coefficients in $\hat{L}(t)$ as well as the independent vector ${\bf{C}}(t)$, are all evaluated only with the instantaneous $\lambda^t$ and $S_z^t$. Hence the corrections ${\bf{\Delta}}(t)$, and consequently the entire dynamics of the electron system, are determined exclusively by  the frozen solutions of the SNLE.
The analytical expressions are the following:
\ba\label{coefM}
&&\hat{L}_{ii}=1+\frac{\hbar S_z^t}{(\hbar^2-{S^t_z}^2)}\!\int\!\frac{d\varepsilon}{\pi}f'(\varepsilon)\mbox{Tr}\!\left[\hat{\rho}^t(\varepsilon)(\varepsilon-\hat{{H}}_t^*)\right]\nonumber\\
&&\hat{L}_{21}=-\int\!\frac{d\varepsilon}{2\pi}f'(\varepsilon)\mbox{Tr}\!\left[\hat{\rho}^t(\varepsilon)\right]\\
&&\hat{L}_{12}=\!-\!\left[\!\frac{\hbar S_z^t}{(\hbar^2-{S^t_z}^2)}\right]^2\!\!\!\int\!\!\frac{d\varepsilon}{\pi}f(\varepsilon)\mbox{Tr}\!\left[\hat{\Gamma}^t\hat{\rho}^t(\varepsilon)(\varepsilon-\hat{{H}}_t^*)\hat{\rho}^t(\varepsilon)\right]\nonumber\\
&&\;\;\;\;\;\;\;\;\;\;\;-\frac{\hbar(\lambda^t-\frac{U}{2})}{S_z^t},\nonumber
\ea
where $\hat{\rho}^t\equiv\sum_\alpha\hat{\rho}_\alpha^0({\bf{V}}_i(t),\lambda^t, S_z^t)$ is the total frozen density matrix and $\hat{H}_t^*\equiv\hat{H}_0^*(\varepsilon^*_i\!\rightarrow\!\varepsilon_i(t)+\lambda^t)$. The effective hybridization reads $\hat{\Gamma}^t=\sum_\alpha\hat{\gamma}_\alpha(t)(1- {S_z^t}^2/\hbar^2)/2$, and $f$ is the Fermi function evaluated at zero bias voltage and temperature difference between the reservoirs. Moreover, 
\ba\label{coefc}
{\bf{C}}_1&=&\frac{S_z^t}{2\left(1-({S^t_z}/\hbar)^2\right)}\!\int\!\!\frac{d\varepsilon}{2\pi}f'(\varepsilon) \frac{d}{dt}\mbox{Tr}\!\!\left[\hat{\rho}^t(\varepsilon){\hat{\Gamma}}^t\right]\\
{\bf{C}}_2&=&\hbar\int\!\!\frac{d\varepsilon}{2\pi}f'(\varepsilon) \mbox{Im}\!\left\{\mbox{Tr}\!\!\left[\hat{G}^{t^{\dagger}}(\varepsilon)\frac{d\hat{G}^t(\varepsilon)}{dt}\hat{\Gamma}^t\right]\right\},\nonumber
\ea
with $\hat{G}^t=[(\varepsilon \hat{I}-\hat{H}_t^*)+i\hat{\Gamma}^t/2]^{-1}$ being the frozen retarded Green's function of the two-level system. We notice that $\hat{L}$ corresponds to the Jacobian matrix of the SNLE evaluated at $\lambda^t$ and $S_z^t$, and therefore it describes steady-state phenomena. On the other hand, 
the time derivatives in ${\bf{C}}$ give rise to terms $\propto\dot{{\bf{V}}}_i$, which describe pumping effects. In particular, the coefficient $C_2$ corresponds to a correction in the number of electrons held by the two level system due to the time-variation of the parameters. 

\begin{figure}[h!]
 \includegraphics[width=0.46\textwidth]{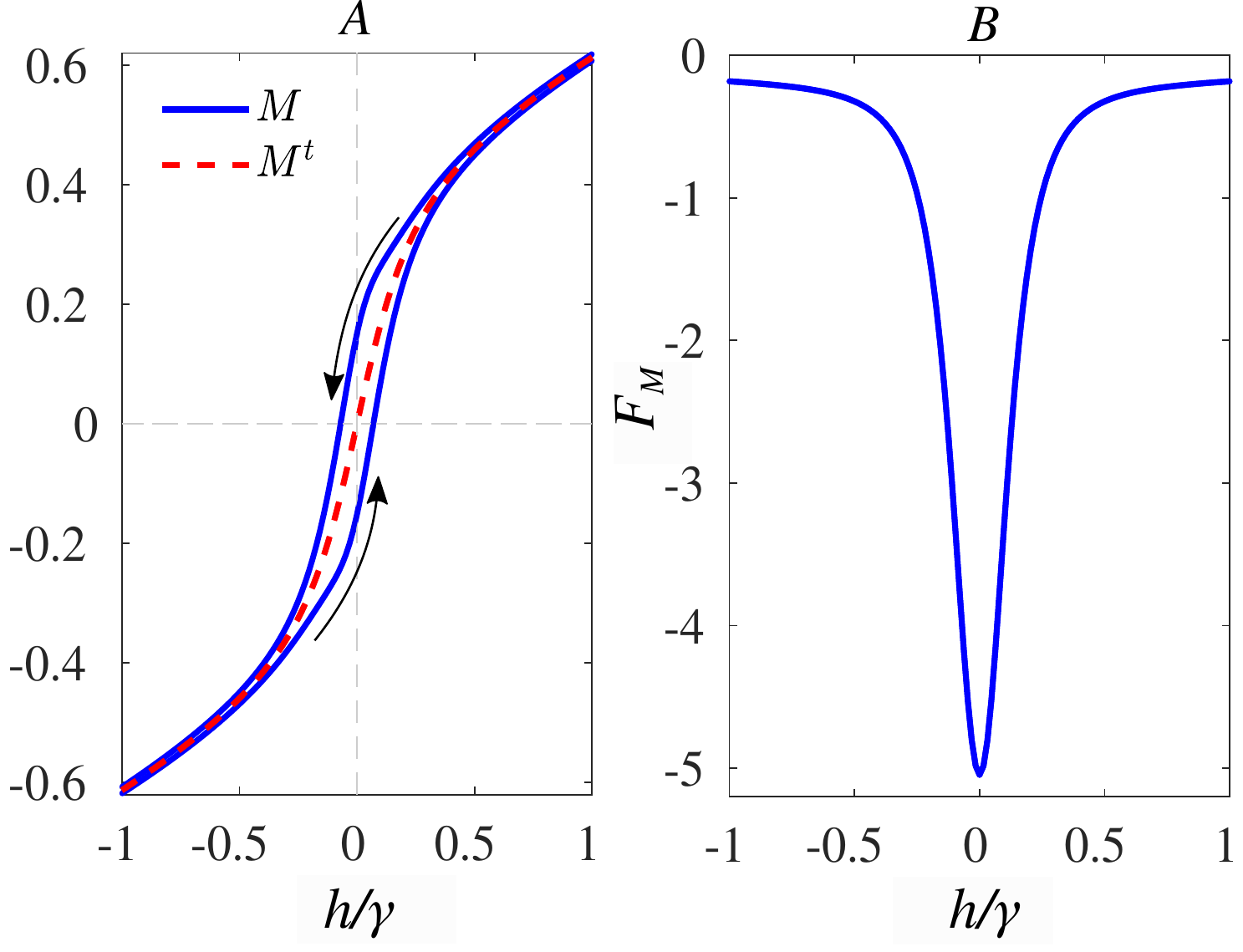}
  \caption{Magnetization of an impurity as a function of the driving Zeeman field $h$. Panel $A$: The blue solid line corresponds to the  magnetization $M$, while the red dashed curve is the frozen value $M^t$. Panel $B$: Pseudo-force $F_M$ in units of $\hbar/\gamma^2$ as a function of $h$. Parameters: $U=4\gamma$, $\hbar\omega=3\times10^{-3}\gamma$, $h_0=3\gamma$, $k_BT=\mu=t_h=0$, and $\varepsilon_\uparrow=-\varepsilon_\downarrow=-h$. Typical experimental values of the hybridization $\gamma\sim 10^{-6}-10^{-5} eV$ correspond to magnetic fields $B\sim10^{-2}-10^{-1}$ T.  }\label{fig2}
\end{figure}

{\it{Example: A magnetically driven Anderson impurity.}}
In this section we show that our description of the adiabatic dynamics is able to capture the main effects of a time-dependent driving. Thus we analyze the behavior of the magnetization $M= n_\uparrow -n_\downarrow$ of a single level Anderson impurity driven by an oscillating Zeeman field $h(t)=h_0\cos(\omega t)$, for $\hbar\omega\ll\gamma$ with $\gamma$ being the total hybridization with the reservoirs. 
The magnetization can be expanded within the adiabatic regime as\cite{ludovico1}
\be M\sim M^t+F_M\,\dot{h},
\ee
where $M^t$ is the frozen magnetization and $F_M$ is a pseudo-force accounting for changes on the magnetization due to the variation of the external field. As reported in Ref. \cite{df} for a constant magnetic field, the instantaneous $M^t$ is a monotonically increasing function of $\vert h \vert$ which vanishes when $h\rightarrow 0$ and reaches its maximum absolute value $M^t\rightarrow \pm1$ at large magnetic fields $\vert h\vert\gg\gamma$ (these features are shown in \cite{supmat}). Nonetheless, the behavior of $M$ departs from the frozen magnetization when $\vert h\vert\lesssim\gamma$, owing to the fact that the effect of the modulation of the field is more perceived within that range. This is shown in Fig. \ref{fig2} -$A$, where we recognize an hysteretic behavior of $M$, in the sense that the impurity is magnetized even when $h=0$. Black arrows indicate the direction in which the curve is traveled during the oscillation, thus we can see that an inversion of the field in required for demagnetization. The retention, or memory, of part of the alignment is due to the presence of a restorative force $F_M\leqslant0$ (see Fig. \ref{fig2} -$B$), which is exclusively originated by the temporal driving. As the field increases $\vert h\vert>\gamma$, the latter force vanishes $F_M\rightarrow 0$, so that $M\rightarrow M^t$. 
It is worth mentioning that the significant increment of the magnetization at zero field $M(h=0)\sim 0.1$ merits a deeper analysis beyond the adiabatic regime.

{\it{Conclusion}}: 
We have developed a mean-field S-S1 technique for describing Coulomb-coupled two level systems in which time-dependent drivings are introduced through the different tunneling elements with the reservoirs $w_{ij}(t)$ as well as by a modulation of the energy levels $\varepsilon_i(t)$. This approach can be applied for quite generic device configurations, in the sense that the finite interaction $U$ can be local as well as non-local, and also the number and connection locations of the reservoirs can be arbitrary. A hopping amplitude between the level (or sites) is contemplated as well. In this way, the S-S1 is capable to describe not only the already well studied AIM but also Coulomb-coupled quantum dots systems where Coulomb-drags effects can take place.

We showed that in the stationary limit but also within the adiabatically driven regime, the effects of the interactions are encoded by only two parameters: a Lagrange multiplier $\lambda$ and the component of the auxiliary spin $S_z$. In the stationary case, the latter are the solutions of the $2\times2$ SNLE in Eqs. (\ref{consteq}) and (\ref{eq12eq}). Particularly, for an adiabatic driving we presented a practical manner to solve the SODE by considering the solutions as little perturbations from the instantaneous (or frozen) values $\lambda^t$ and $S_z^t$, which are found by solving again the SNLE but at every instant of time. Then, the corresponding perturbative corrections $\Delta\lambda (\lambda^t,S_z^t)$ and $\Delta S_z(\lambda^t,S_z^t)$ are simply the solutions of a SLE in Eq. (\ref{cor}) evaluated only with the frozen values. Finally we considered a magnetically driven Anderson impurity as an example to show that the above perturbative treatment, in spite of being simple, it is capable to predict interesting phenomena of a pure dynamical nature. More precisely we found a finite magnetization of the impurity even at zero magnetic field, which is merely originated by the time-dependent driving. 

{\it{Acknowledgments}}: We acknowledge support from H2020 Framework Programme, under ERC Advanced Grant No. 692670 “FIRSTORM” and  from Ministero dell’Istruzione, dell’Università e della Ricerca under PRIN 2015 (Prot. 2015C5SEJJ001) and PRIN 2017 "CEnTraL". We also thank D. Guerci and L. Fanfarillo for useful discussions.



\newpage
\setcounter{figure}{0}
\setcounter{equation}{0}
\renewcommand{\thefigure}{S\arabic{figure}}  
\renewcommand{\theequation}{S\arabic{equation}} 
\renewcommand{\figurename}{Figure} 
\renewcommand{\thefootnote}{S\arabic{reference}}

\pagebreak
\onecolumngrid
\pagebreak

\section*{Slave-Spin 1 formulation: A simple approach to 
time-dependent transport through an interacting two level system. Supplemental Material}

\subsection*{Spin expectation values}
From the general norm one spin state in Eq. (9), it is possible to compute the expectation value of any spin operator $\langle \mathcal{O}_S\rangle_s=g_{\mathcal{O}_S}(d,e)$ as a function of the double occupied and empty amplitudes. In particular, 
\be
\begin{matrix}\label{s+}
    \langle S^+\rangle_s=\langle S_x\rangle_s+i\langle S_y\rangle_s=\frac{2\hbar}{\sqrt{2}}s(d^*+e),\,\,\,\,\,\, & \langle \{S_z,S^+\}\rangle_s=\frac{2\hbar^2}{\sqrt{2}}s(d^*-e)
    \end{matrix}
\ee
\be
\begin{matrix}
    \langle S_z\rangle_s=\hbar(\vert d\vert^2-\vert e\vert^2),\,\,\, & \langle S_z^2\rangle_s =\hbar^2(\vert d\vert^2+\vert e\vert^2),\,\,\, &  \langle{S^+}^2\rangle_s=2\hbar^2 d^*e,\nonumber
\end{matrix}
\ee
with $s=\sqrt{1-(\vert d\vert^2+\vert e\vert^2)}$ being the amplitude of the $\vert S_z=0\rangle$ state.
\subsection*{Stationary limit}
By taking the real part of Eq. (7) multiplied by $\langle S^-\rangle_s$, we get
\be\label{set2a}
\hbar\dot{\phi}^+\vert\langle S^+\rangle_s\vert^2=(\frac{U}{2}-\lambda)\vert\langle S^+\rangle_s\vert^2+\hbar U\langle S_z\rangle_s\left(1-\frac{\langle S_z^2\rangle_s}{\hbar^2}\right)
-2\,\hbar\langle S_z\rangle_s\,\sum_{i={1,2}}\sum_{\alpha,k_{\alpha}}\mbox{Re}\{\tilde{w}_{\alpha,i} \langle c^\dagger_{k_\alpha}d^*_i\rangle_f\},
\ee
where $\phi^+$ is the complex argument of $\langle S^+\rangle_s=\vert\langle S^+\rangle_s\vert e^{i\phi^+}$ when written in a phasor form. On the other hand, the imaginary part reads 
\be\label{set2b}
d_t \vert\langle S^+\rangle_s\vert^2=4\hbar Us^2\mbox{Im}\{d\,e\}-d_t\langle S_z\rangle_s^2,
\ee
or equivalently  
\vspace{-4mm}
\be\label{set2bb}
\vert\langle S^+\rangle_s\vert^2(t)=\beta-\langle S_z\rangle_s^2(t)+4\hbar U\int^t_{t_0}dt's^2(t')\mbox{Im}\{d(t')e(t')\},
\ee
with $\beta$ being a constant number.
In the stationary case, all the spin expectation values remain constant $d_t\langle{\mathcal{O}}_S\rangle_s^0=0$, so that from Eq. (\ref{set2b}) we find
\be
 {\mbox{Im}\{d\,e\}=0} \Leftrightarrow
  \begin{cases}
    d,\, e       & \quad \in \mathbb{R} \\
    \phi_d=-\phi_e+m\pi  & \quad \text{with } m \in  \mathbb{Z},
  \end{cases}
\ee
where $\phi_l$ are the phases of the complex amplitudes $l=\vert l\vert e^{i\phi_l}$, with $l=d,e$.
The above possibilities for $d$ and $e$, are sort of different ``Gauge choices". Now, by plugging the latter condition into Eq. (\ref{set2bb}) we can see that in the stationary limit $\vert\langle S^+\rangle^0_s\vert^2=\beta -{\langle S_z\rangle_s^0}^2$. The constant $\beta$, can be obtained from the partial derivatives
\be
\frac{\partial \vert\langle S^+ \rangle_s^0\vert^2}{\partial \vert d\vert^2}=-2\hbar^2\langle S_z\rangle_s^0=-\frac{\partial \vert\langle S^+ \rangle_s^0\vert^2}{\partial \vert e\vert^2},
\ee
which is satisfied for $2\vert d\vert\vert e\vert=1-\vert d\vert^2-\vert e\vert^2$, and leads to $\beta=\hbar^2$ and $\langle S_z^2\rangle_s^0=({\langle S_z\rangle_s^0}^2+\hbar^2)/{2}$ when 
replaced into Eq. (\ref{s+}).

On the other hand, the stationary condition also sets $\dot{\phi^+}=0$, so that Eq. (\ref{set2a}) reads
\be\label{set3}
0=\left[\lambda^0 -\frac{U}{2}\left(1+\frac{\langle S_z\rangle_s^0}{\hbar}\right)\right]\left(1-\frac{{\langle S_z\rangle_s^0}^2}{\hbar^2}\right)+2\frac{\langle S_z\rangle_s^0}{\hbar}\,\sum_{i={1,2}}\sum_{\alpha,k_{\alpha}}\mbox{Re}\{\tilde{w}_{\alpha,i} \langle c^\dagger_{k_\alpha}d_i^*\rangle_f\}.
\ee
The same is obtained when starting from Eq. (8).

Finally, by following Ref. \cite{sup_jauho} we find the final expressions of the SNLE in Eq. (10) and (11) of the main text, where the partial density matrix reads $\hat{\rho}_\alpha^0(\varepsilon)=\hat{G}^0(\varepsilon)\hat{\Gamma}_\alpha^0\hat{G}^0(\varepsilon)$, with $\hat{G}^0=[(\varepsilon\hat{I}-\hat{H}^*_0)+i\sum_\alpha\hat{\Gamma}^0_\alpha/2]^{-1}$ being the retarded Green's function of the two-level subsystem when connected to the reservoirs, and $\hat{\Gamma}_\alpha=\hat{\gamma}_\alpha \vert\langle S^+\rangle^0_s\vert^2/(2\hbar^2)$ is the renormalized hybridization with the $\alpha$-reservoir due to the interactions. The bare hybridization matrix reads $[\hat{\gamma}_\alpha]_{ij}=w_{\alpha,i}w_{\alpha,j}\varrho_\alpha $ where $\varrho_\alpha$ corresponds to the density of state of the $\alpha$-lead which is considered energy-independent. 
\vspace{-8mm}
\subsection*{Comparison with Kotliar-Ruckenstein predictions}
Here we show the results given by the {\it{stationary}} S-S1 formulation when applied to describe a single level magnetic impurity, and we also compare with Kotliar-Ruckenstein (K-R) predictions \cite{sup_kr,sup_dong}. In particular, we exhibit the behavior of the electrical conductance $G$
and the spin amplitudes $d$, $e$ as functions of the energy level of the impurity $\varepsilon_0$.

\begin{figure}[!ht]
\centering
  \includegraphics[width=0.5\textwidth]{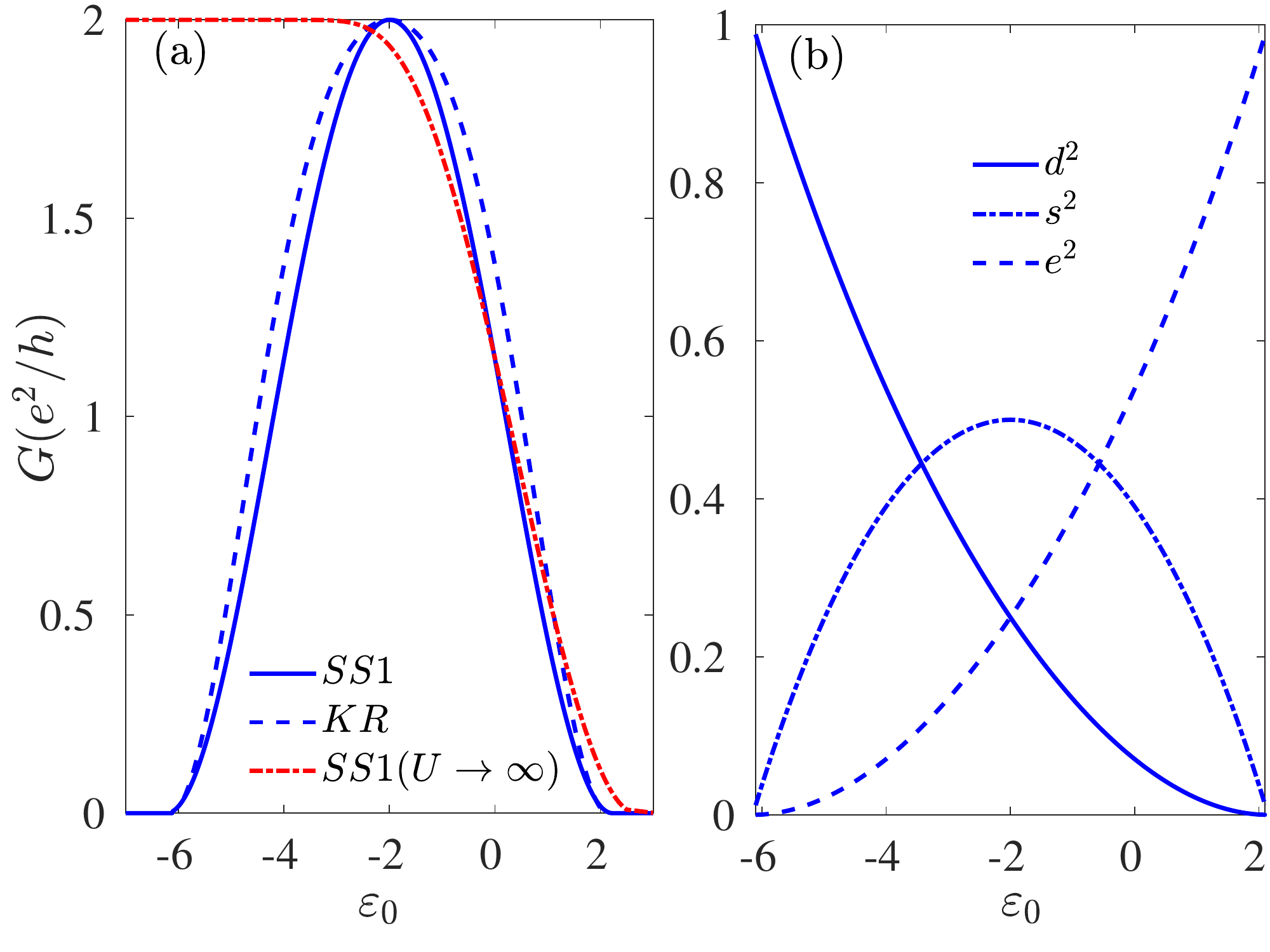}
  \caption{Panel (a): Electrical conductance $G$ as a function of the energy level of the impurity $\varepsilon_\uparrow=\varepsilon_\downarrow=\varepsilon_0$. S-S1 predictions are shown for $U=4\gamma$ (blue solid line) with $\gamma$ being the total hybridization with the reservoirs, as well as in the strongly interacting limit $U\rightarrow \infty$ (Red dot-dashed curve). Finally, the blue-dashed line corresponds to the K-R predictions in the case of $U=4\gamma$. Panel (b): Behavior of $d^2$, $e^2$ and $s^2=1-(d^2+e^2)$ as a function of $\varepsilon_0$. Other parameters are $k_BT=\mu=t_h=0$}\label{des}
\end{figure}

From the above figure we can see that the S-S1 approach reproduces well the peak in the linear conductance $G=2e^{2}/h$ at the symmetric point $\varepsilon_0=-U/2$, as well as the critical energies in which the conductance is significantly suppressed (around $\varepsilon_0\sim -6\gamma$ and $\varepsilon_0\sim 2\gamma$, with $\gamma$ being the total hybridization with the reservoirs). For intermediate energies, the values of $G$ slightly differ from the K-R curve, which is not worrying since both of the formulations provide qualitative predictions within this energy range. On the other hand, in the strongly interacting limit $U\rightarrow\infty$ the conductance shows a strong enhancement from energies a bit below the Fermi level of the reservoirs $\varepsilon_0\lesssim 0$ and reaches its maximum for $-2\gamma>\varepsilon_0>-3\gamma$, as it was experimentally observed \cite{sup_dong,sup_exp}. 

Then, in panel (b) we depict the amplitudes of the double occupied $d^2$ and empty $e^2$ states, and the corresponding $s^2=1-(d^2+e^2)$ for $U=4\gamma$. As expected, we can see that when the energy is far bellow the Fermi level $\varepsilon_0\ll 0$, then the impurity is double occupied so that $d^2=1$ and $e^2=0$.
As the energy level $\varepsilon_0$ goes up, $d^2$ inevitably decreases and the
singly occupied state amplitude $s^2$ starts to increase and reaches its maximum at the symmetric point $\varepsilon_0=-U/2$. After that, with a further increase of the energy level, there are no electrons in the impurity so that $e^2=1$ and $d^2=0$.
\subsection*{An adiabatic driving}
According to Eq. (12), we proceed to evaluate the SODE up to first order in the small rate of change of the time-dependent parameters $\dot{{\bf{V}}}_i$. For that, we start from Eq. (\ref{set2b}) in which we approximate $d_t\vert\langle S^+\rangle_s\vert^2\sim d_t\vert {S^+}^t\vert^2$ and $d_t\langle S_z\rangle_s^2\sim d_t  {S_z^t}^2$ so that
\vspace{-3mm}
\be\label{set2bad}
d_t\vert {S^+}^t\vert^2=4\hbar Us^2\mbox{Im}\{d\,e\}-d_t  {S_z^t}^2.
\ee
Now, since the frozen values satisfy the relation $\vert {S^+}^t\vert^2=\hbar^2 -{S_z^t}^2$, we get again that $\mbox{Im}\{d\,e\}=0$ $\forall t$ as long as the evolution of the system is quasi-static. In the same way as for the stationary case, the latter condition leads to $\vert\langle S^+\rangle^{ad}_s\vert^2=\hbar^2 -{\langle S_z\rangle_s^{ad}}^2$ and $\langle S_z^2\rangle_s^{ad}=({\langle S_z\rangle_s^{ad}}^2+\hbar^2)/{2}$.

On the other hand, when writing down Eq. (\ref{set2a}) up to first order, we should also approximate $\dot{\phi^+}\sim \dot{{\phi^+}}^f$. The phase ${\phi^+}^f$ behaves as a Gauge choice due to its static nature, and therefore it does not depend on time (or equivalently, on the parameters of the system) so that we have $\dot{{\phi^+}}^f=0$. 
Finally, by following Ref. \cite{sup_ludovico-capone} we can perform the following approximations in Eqs. (6) and (\ref{set2a}) 
\ba
\sum_{i={1,2}}\sum_{k_{\alpha}}2\mbox{Re}\{\tilde{w}_{\alpha,i} \langle c^\dagger_{k_\alpha}d^*_i\rangle_f\}&\sim&\int\!\!\frac{d\varepsilon}{2\pi}\left(\mbox{Re}\!\left\{\!\mbox{Tr}\!\!\left[\hat{G}^t(\varepsilon){\hat{\Gamma}}_\alpha(t)\right]\!\right\}\!f_\alpha(\varepsilon)-\frac{\hbar}{2} f'(\varepsilon) \frac{d}{dt}\mbox{Tr}\!\!\left[\hat{\rho}_\alpha^t(\varepsilon){\hat{\Gamma}}(t)\right]\right),
\ea
\ba\label{constad}
\frac{\langle S_z\rangle_s^{ad}}{\hbar}+1=\langle n_{\mathcal{T}}^*\rangle_f&\sim&\int\!\frac{d\varepsilon}{2\pi}\Bigg(\sum_\alpha\!\mbox{Tr}\left[\hat{\rho}_\alpha^t(\varepsilon)\right]f_\alpha(\varepsilon)
+\hbar f'(\varepsilon) \mbox{Im}\!\left\{\mbox{Tr}\!\!\left[{\hat{G}}^{t^{\dagger}}(\varepsilon)\frac{d\hat{G}^t(\varepsilon)}{dt}\hat{\Gamma}(t)\right]\right\}\Bigg).
\ea
In this way, the linear response evaluation of Eq. (\ref{set2a}) reads
\ba\label{set2aad}
0&=&\left[\lambda^{ad}(t)-\frac{U}{2}\left(1+\frac{\langle S_z\rangle_s^{ad}(t)}{\hbar}\right)\right]\left(1-\frac{{\langle S_z\rangle_s^{ad}}^2(t)}{\hbar^2}\right)+\frac{\langle S_z\rangle_s^{ad}(t)}{\hbar}\!\int\!\!\frac{d\varepsilon}{2\pi}\left(\sum_{\mathclap{\alpha}}\mbox{Re}\!\left\{\!\mbox{Tr}\!\!\left[\hat{G}^t(\varepsilon){\hat{\Gamma}}_\alpha(t)\right]\!\right\}\!f_\alpha(\varepsilon)\right.\\
&&-\frac{\hbar}{2} f'(\varepsilon) \frac{d}{dt}\mbox{Tr}\!\!\left[\hat{\rho}^t(\varepsilon){\hat{\Gamma}}(t)\right]\Bigg),\nonumber
\ea
hence, we were able to reduce the SODE to a system of non-linear equations composed by Eqs. (\ref{constad}) and (\ref{set2aad}).
\vspace{-4mm}
\section{Behavior of $M^t$}
\vspace{-8mm}
\begin{figure}[h!]
\centering
  \includegraphics[width=0.44\textwidth]{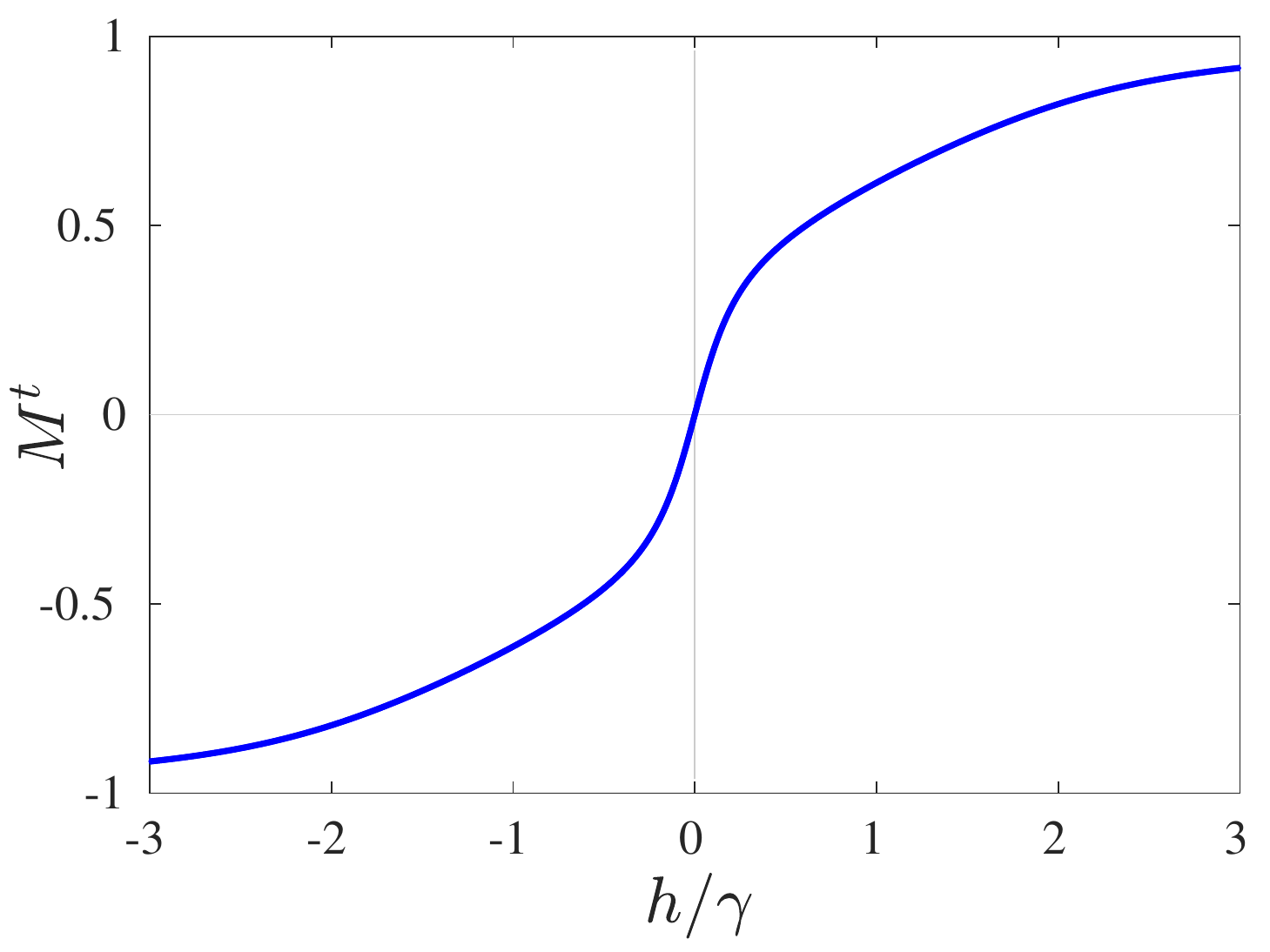}
  \caption{Frozen magnetization $M^t$ as a function of the Zeeman field $h$, which is expressed in units of the total bare hybridization with the reservoirs $\gamma$. All the parameters are the same as in Fig. 2.}\label{mt}
\end{figure}


\begin{thebibliography}{99}



\bibitem{kondo}J. Kondo, Resistance Minimum in Dilute Magnetic Alloys, Prog. Theor. Phys. 32, 37 (1964).


\bibitem{dragor} M. B. Pogrebinskii, Fiz. Tekh. Poluprovodn. 11, 637 (1977) (Sov. Phys. Semicond.11 , 372 (1977)).

\bibitem{dragreview} B. N. Narozhny and A. Levchenko, Coulomb Drag, Rev. Mod. Phys. 88, 025003 (2016).


\bibitem{dragsa}D. Venturelli, R. Fazio, and V. Giovannetti, Minimal Self-Contained Quantum Refrigeration Machine Based on Four Quantum Dots, Phys. Rev. Lett. 110, 256801 (2013). 

\bibitem{dragst5} T. Ruokola and T. Ojanen, Single-electron heat diode: Asymmetric heat transport between electronic reservoirs through Coulomb islands, Phys. Rev. B 83, 241404(R) (2011).

\bibitem{dragst1} R. S\'anchez, R. L\'opez, D.S\'anchez, and M.   B\"uttiker, Mesoscopic Coulomb Drag, Broken Detailed Balance, and Fluctuation Relations, Phys. Rev. Lett. 104, 076801 (2010).

\bibitem{dragst2} J. Soo Lim, R. L\'opez, and D. S\'anchez, Engineering drag currents in Coulomb coupled quantum dots, New J. Phys. 20, 023038 (2018).

\bibitem{dragst3} Y. Zhang, G. Lin, and J. Chen, Three-terminal quantum-dot refrigerators, Phys. Rev. E 91, 052118 (2015).

\bibitem{dragst4} B. Bhandari, G. Chiriac\`o, P. A. Erdman, R. Fazio, and F. Taddei, Thermal drag in electronic conductors, Phys. Rev. B 98, 035415 (2018).


\bibitem{dragst6} M. A. Sierra, D. S\'anchez, A. P. Jauho, K. Kaasbjerg, Fluctuation-driven Coulomb drag in interacting quantum dot systems, arXiv:1903.02996v1.



\bibitem{dragse1} A. J. Keller, J. S. Lim, D. S\'anchez, R. L\'opez, S. Amasha, J. A. Katine, H. Shtrikman, and D. Goldhaber-Gordon, Cotunneling Drag Effect in Coulomb-Coupled Quantum Dots, Phys. Rev. Lett. 117, 066602 (2016).

\bibitem{dragse2} G. Shinkai, T. Hayashi, T. Ota, K. Muraki, and T. Fujisawa, Bidirectional Current Drag Induced by Two-Electron Cotunneling in Coupled Double Quantum Dots, Appl. Phys. Express 2, 081101 (2009).





\bibitem{ludovico1} M. F. Ludovico, F. Battista, F. von Oppen, and L. Arrachea, Adiabatic response and quantum thermoelectrics for ac-driven
quantum systems, Phys.Rev.B 93, 075136 (2016).

\bibitem{thierschmann}H. Thierschmann, R. S\'anchez, B. Sothmann,  H. Buhmann, and  L. W. Molenkamp, Thermoelectrics  with
Coulomb-coupled  quantum  dots, C. R. Physique 17, 1109 (2016).

\bibitem{erdman}P. A. Erdman, V. Cavina, R. Fazio, F. Taddei, V. Giovannetti, Maximum power and corresponding efficiency for two-level quantum heat engines and refrigerators, arXiv preprint arXiv:1812.05089.

\bibitem{lim} J. S. Lim, R. L\'opez, and D. S\'anchez, Dynamic thermoelectric and heat transport in mesoscopic capacitors,
Phys.Rev.B 88, 201304(R) (2013).

\bibitem{brandner} K. Brandner, K. Saito, and U. Seifert, Thermodynamics of micro and nano systems driven by periodic temperature variations, Phys. Rev. X 5, 031019 (2015).


\bibitem{thierschmann2} H. Thierschmann, R. S\'anchez, B. Sothmann, F. Arnold, C. Heyn, W. Hansen, H. Buhmann, and L. W. Molenkamp, Three-terminal energy harvester with coupled quantum dots,” Nature Nanotech. 10, 854 (2015).

\bibitem{ludovico2}M. F. Ludovico and M. Capone, Enhanced performance of a quantum-dot-based nanomotor due to Coulomb interactions, Phys. Rev. B 98, 235409 (2018).


\bibitem{qo1} E. Bocquillon, F. D. Parmentier, C. Grenier, J.-M. Berroir, P. Degiovanni, D. C. Glattli, B. Pla\c{c}ais, A. Cavanna, Y. Jin, and G. F\`eve, Electron quantum optics: partitioning electrons one by one, Phys. Rev. Lett. 108, 196803 (2012).

\bibitem{qo2} J. Dubois, T. Jullien, F. Portier, P. Roche, A. Cavanna, Y. Jin, W. Wegscheider, P. Roulleau, and D. C. Glattli, Minimal-excitation states for electron quantum
optics using levitons, Nature 502, 659–663 (2013).


\bibitem{romero}J. I. Romero, P. Roura-Bas, A. A. Aligia, and L. Arrachea, Nonlinear charge and energy dynamics of an adiabatically driven interacting quantum dot, Phys. Rev. B 95, 235117 (2017).

\bibitem{anders}F. B. Anders and A. Schiller, Spin precession and real-time dynamics in the Kondo model: Time-dependent numerical renormalization-group study, Phys. Rev. B 74, 245113 (2006).

\bibitem{schiro}M. Schir\'o and M. Fabrizio, Real-time diagrammatic Monte Carlo for nonequilibrium quantum transport, Phys. Rev. B 79, 153302 (2009).

\bibitem{schiro2} M. Schir\'o and M. Fabrizio,Time-Dependent Mean Field Theory for Quench Dynamics in Correlated Electron Systems, Phys. Rev. Lett. 105, 076401 (2010).



\bibitem{coleman} P. Coleman, New approach to the mixed-valence problem, Phys. Rev. B 29, 3035 (1984).

\bibitem{KR} G. Kotliar and A. E. Ruckenstein, New functional integral approach to strongly correlated Fermi systems: The Gutzwiller approximation as a saddle point, Phys. Rev. Lett. 57, 1362 (1986).

\bibitem{demedici}L. de'Medici, A. Georges, and S. Biermann, Orbital-selective Mott transition in multiband systems: Slave-spin representation and dynamical mean-field theory, Phys.Rev.B 72, 205124 (2005).

\bibitem{huber}A. R\"uegg, S. D. Huber and M. Sigrist, Z2-slave-spin theory for strongly correlated fermions, Phys. Rev. B 81, 155118 (2010).

\bibitem{df} D. Guerci and M. Fabrizio, Unbinding slave spins in the Anderson impurity model, Phys. Rev. B 96, 201106(R) (2017).

\bibitem{caponedemedici} L. de’ Medici and M. Capone, Modeling many-body
physics with slave-spin mean-field: Mott and Hund’s physics in fe-superconductors, in The Iron Pnictide Super- conductors: An Introduction and Overview, edited by F. Mancini and R. Citro (Springer International Publishing, Cham, 2017), pp. 115–185


\bibitem{citro}R. Citro and F. Romeo, Non-equilibrium slave bosons approach to quantum pumping in interacting quantum dots, J.  Phys.: Conf. Ser. 696, 012014 (2016).
\bibitem{dong}B. Dong, G. H. Ding and X. L. Lei, Time-dependent quantum transport through an interacting quantum dot beyond sequential tunneling: second-order quantum rate equations, J. Phys.: Condens. Matter 27, 205303 (2015).  
\bibitem{daniele}D. Guerci, Transport through a magnetic impurity: a slave-spin approach, Phys. Rev. B 99, 195409 (2019).

\bibitem{supmat} Supplemental material

\bibitem{footnote}Without loss of generality, and due to the fact that global phase factors are irrelevant for quantum states, we have chosen the first coefficient to be a real number while $d$ and $e$ remain being complex.

\end{thebibliography}

\begin{thebibliography}{99}
\bibitem{sup_jauho}
A. P. Jauho, Introduction to the Keldysh nonequilibrium Green function technique.
Lecture notes (2006).
\bibitem{sup_kr}G. Kotliar and A. E. Ruckenstein, New functional integral approach to strongly correlated Fermi systems: The Gutzwiller approximation as a saddle point, Phys. Rev. Lett. 57, 1362 (1986).
\bibitem{sup_dong} B. Dong and X. L. Lei, Kondo-type transport through a quantum dot under magnetic fields, Phys. Rev. B 63, 235306 (2001).
\bibitem{sup_exp} S. M. Cronenwett, T. H. Oosterkamp, L. P. Kouwenhoven, A Tunable Kondo Effect in Quantum Dots, Science 281, 540 (1998).
\bibitem{sup_ludovico-capone}M. F. Ludovico and M. Capone, Enhanced performance of a quantum-dot-based nanomotor due to Coulomb interactions, Phys. Rev. B 98, 235409 (2018). 

\end{thebibliography}
\end{document}